\documentclass[preprint,showpacs,preprintnumbers,amssymb,aps,nofootinbib]{revtex4}
\usepackage{amsfonts,color,graphicx,epsfig,graphicx,amsmath,multirow,slashed}

\begin{document}
\preprint{PITT PACC-1908}
\title{Inclusive productions of $\Upsilon(1S,2S,3S)$ and $\chi_b(1P,2P,3P)$ via the Higgs boson decay}
\author{Zhan Sun$^1$}
\email{zhansun@cqu.edu.cn}
\author{Yang Ma$^2$}
\email{mayangluon@pitt.edu}
\affiliation{
\footnotesize
$^{1}$ Department of Physics, Guizhou Minzu University, Guiyang 550025, People's Republic of China \\
$^{2}$ PITT-PACC, Department of Physics and Astronomy, University of Pittsburgh, PA 15260, USA
}

\date{\today}

\begin{abstract}
In this paper, we carry out the complete $\mathcal O(\alpha\alpha_s^{2})$-order study on the inclusive productions of $\Upsilon(nS)$ and $\chi_b(nP)$ ($n=1,2,3$) via the Standard Model Higgs boson decay,  within the framework of nonrelativistic QCD. The feeddown effects via the higher excited states are found to be substantial. The color-octet $^3S_1^{[8]}$ state related processes consisting of $H^0 \to b\bar{b}[^3S_1^{[8]}]+g$ and $H^0 \to b\bar{b}[^3S_1^{[8]}]+Q+\bar{Q}$ ($Q=c,b$) play a vital role in the predictions on the decay widths. Moreover, our newly calculated next-to-leading order QCD corrections to $H^0 \to b\bar{b}[^3S_1^{[8]}]+g$ can enhance its leading-order result by 3-4 times, subsequently magnifying the total $^3S_1^{[8]}$ contributions by about $40\%$. Such a remarkable enhancement will to a large extent influence the phenomenological conclusions. For the color-singlet $^3P_J^{[1]}$ state, in addition to $H^{0} \to b\bar{b}[^3P_J^{[1]}]+b+\bar{b}$, the newly introduced light hadrons associated process, $H^{0} \to b\bar{b}[^3P_J^{[1]}]+g+g$, can also provide non-negligible contributions, especially for $^3P_2^{[1]}$. Summing up all the contributions, we have $\mathcal B_{H^0 \to \chi_b(nP)+X} \sim 10^{-6}-10^{-5}$ and $\mathcal B_{H^0 \to \Upsilon(nS)+X} \sim 10^{-5}-10^{-4}$, which meets marginally nowadays LHC experimental data and can help in understanding the heavy quarkonium production mechanism as well as the Yukawa couplings.
\pacs{12.38.Bx, 12.39.Jh, 14.40.Pq}

\end{abstract}

\maketitle

\section{Introduction}
Bottomonium, as the heaviest bound state, has its own advantages comparing to the charmonium. Due to the large mass of the constituent heavy quarks, both its typical coupling constant $\alpha_s$ and relative velocity $v$ are smaller than those of charmonium. As a result, the perturbative results over the expansion of $\alpha_s$ and $v^2$ for bottomonium will be more convergent than the charmonium case, which makes $b\bar{b}$ mesons an even better place to apply the nonrelativistic QCD (NRQCD) framework \cite{Bodwin:1994jh}. 

Among the bottomonium family, the $\Upsilon$ and $\chi_b$ are most studied because the two mesons can be easily detected by hunting their decaying into lepton pairs\footnote{The decay of $\chi_b$ into lepton pair is indirect, $\chi_b \to \Upsilon+\gamma \to l^+l^-+\gamma$.}. Earlier studies of $\Upsilon$ and $\chi_b$ productions can be found in Refs. \cite{Braaten:2000gw,Artoisenet:2008fc,Wang:2012is,Likhoded:2012hw,Gong:2013qka,Han:2014kxa,Feng:2015wka,Sun:2013wuk,Sun:2014kva} and references therein, where the NRQCD predictions succeeded in explaining almost all the existing experimental measurements. However, considering the fact that the color-octet (CO) long distance matrix elements (LDMEs) that used to well explain the hadroprodution of $J/\psi$ leads to dramatic discrepancies between the theoretical predictions and the measured total cross sections of $e^+e^- \to J/\psi+X_{\textrm{non}-c\bar{c}}$ from the $BABAR$ and Belle collaborations \cite{Zhang:2009ym}, it is indispensable to take investigations on the $\Upsilon(nS)$ and $\chi_b(nP)$ productions in a variety of other processes to further test the validity and universality of the CO LDMEs.

The Higgs boson decay provides a good chance for the studies on $\Upsilon$ and $\chi_b$ because of the large number of $H^0$ events at the high energy colliders, e.g., the HL-LHC and HE-LHC can produce $1.65 \times 10^8$ and $5.78 \times 10^8$ $H^0$ events each year, respectively \cite{Jiang:2015pah}. Although the number of $H^0$ events at the Circular Electron Positron Collider (CEPC) can only reach up to $1.1 \times 10^6$ per year \cite{Jiang:2015pah,Qiao:1998kv,Liao:2018nab}, the ``clean" background of CEPC comparing to LHC may help us to more easily hunt the heavy quarkonium related processes. Pioneering studies of inclusive $\Upsilon$ and $\chi_b$ productions via $H^0$ decay can be found in Refs. \cite{Jiang:2015pah,Qiao:1998kv,Liao:2018nab}. Qiao et al. studied the direct (no feeddown contributions) inclusive production of $\Upsilon(1S)$ via  $H^0$ decay, including both color-singlet (CS) and CO contributions \cite{Qiao:1998kv}. Based on the CS mechanism, the investigations on the semi-inclusive productions of $\Upsilon$ and $\chi_b$ in association with a $b\bar{b}$ pair, $H^{0} \to b\bar{b}[^3S_1^{[1]},^3P_J^{[1]}]+b+\bar{b}$, were carried out by Liao et al. \cite{Liao:2018nab}. Note that, in addition to the processes in \cite{Liao:2018nab}, the other CS process, $H^{0} \to b\bar{b}[^3P_J^{[1]}]+g+g$, might also have remarkable contributions to $\chi_b$ production. Moreover, we learned from the inclusive productions of heavy quarkonium via the $Z$ boson decay that the $^3S_1^{[8]}$ state played a vital role. As shown in our recent work \cite{Sun:2018hpb}, the lowest order process of the $^3S_1^{[8]}$ state, $Z \to Q\bar{Q}[^3S_1^{[8]}]+g$, could receive a remarkable positive NLO QCD correction, which considerably enhance the NRQCD predictions. It is then natural to wonder whether the NLO QCD corrections to $H^0 \to b\bar{b}[^3S_1^{[8]}]+g$ can bring a similar significant enhancement on the LO results, so as to influence the phenomenological conclusions markedly. Besides the vital sense in the studies on the production mechanism of the heavy quarkonium, the decay of the Higgs boson into heavy quarkonium is also very helpful for understanding the electroweak breaking mechanism, especially the Yukawa couplings. In view of these points, we use NRQCD to have a complete $\mathcal O(\alpha\alpha_s^{2})$-order analysis on the inclusive productions of $\Upsilon(1S,2S,3S)$ and $\chi_b(1P,2P,3P)$ via $H^0$ decay, where all necessary feeddown effects are included.

The rest of the paper is organized as follows: In Sec. II, we give a description on the calculation formalism. In Sec. III, the phenomenological results and discussions are presented. Section IV is reserved as a summary.

\section{Calculation Formalism}

\begin{figure*}
\includegraphics[width=0.95\textwidth]{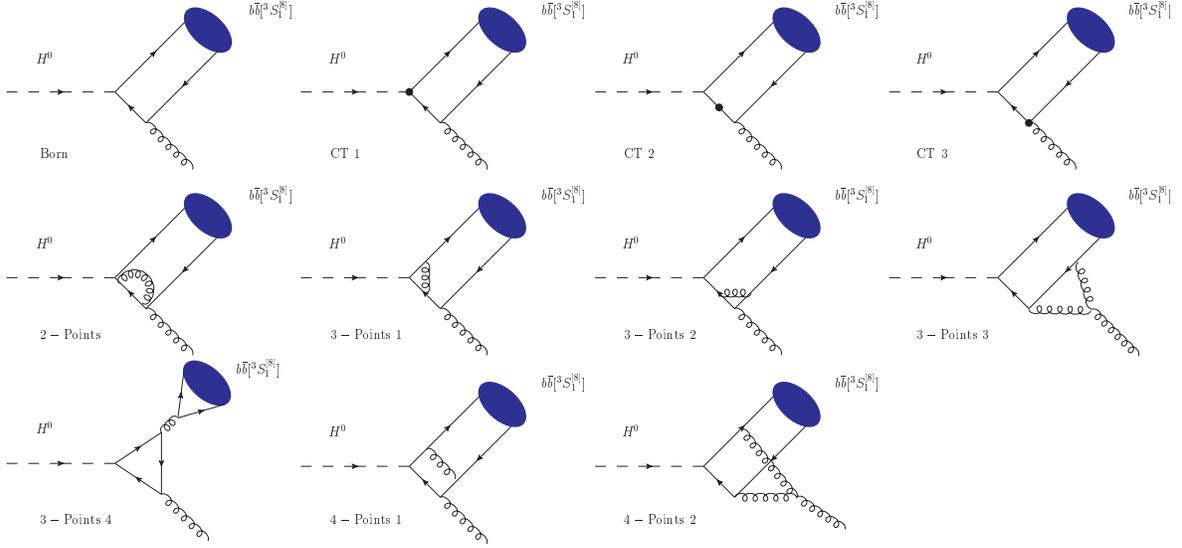}
\caption{\label{fig:Feyn1}
Typical Feynman diagrams for the virtual corrections to the process of $H^0 \to b\bar{b}[^3S_1^{[8]}]+g$. The superscript ``CT" denotes the counterterms.}
\end{figure*}

\begin{figure*}
\includegraphics[width=0.95\textwidth]{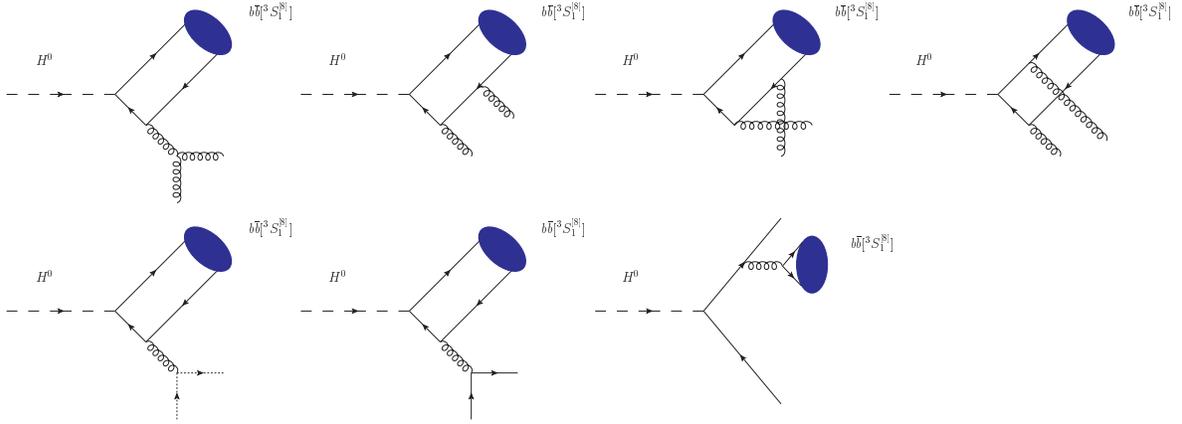}
\caption{\label{fig:Feyn2}
Typical Feynman diagrams for the real corrections to the process of $H^0 \to b\bar{b}[^3S_1^{[8]}]+g$.}
\end{figure*}

\begin{figure*}
\includegraphics[width=0.75\textwidth]{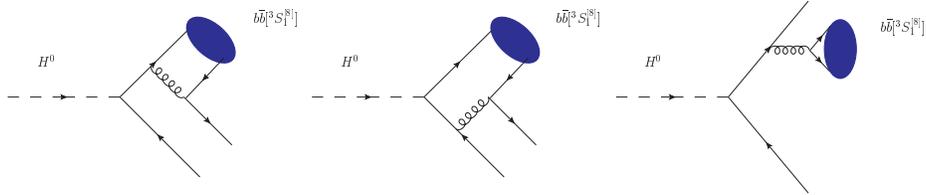}
\caption{\label{fig:Feyn3}
Typical Feynman diagrams for the $\textrm{NLO}^{*}$ processes of $^3S_1^{[8]}$, $H^0 \to b\bar{b}[^3S_1^{[8]}]+Q+\bar{Q}$, where $Q=c,b$. For $Q=c$, the first two diagrams are excluded.}
\end{figure*}

Within the NRQCD framework, the decay width of $H^0 \to \Upsilon(\chi_b)+X$ can be written as:
\begin{eqnarray}
d\Gamma=\sum_{n}d\hat{\Gamma}_{n}\langle \mathcal O ^{H}(n)\rangle,
\end{eqnarray}
where $d\hat{\Gamma}_n$ is the perturbative calculable short distance coefficients (SDCs), representing the production of a configuration of the $Q\bar{Q}$ intermediate state with a quantum number $n(^{2S+1}L_J^{[1,8]})$. $\langle \mathcal O ^{H}(n)\rangle$ is the universal nonperturbative LDME. At LO accuracy in $v$, for the $\Upsilon$ case, four states should be included, i.e. $b\bar{b}[^3S_{1}^{[1]}],b\bar{b}[^1S_{0}^{[8]}],b\bar{b}[^3S_{1}^{[8]}]$, and $b\bar{b}[^3P_{J}^{[8]}]$. While in the case of $\chi_b$, we only need consider $b\bar{b}[^3S_1^{[8]}]$ and $b\bar{b}[^3P_J^{[1]}]$. All the involved processes are listed below:
\begin{itemize}
\item
For $n=^3S_1^{[8]}$, up to $\mathcal O(\alpha_s^2)$ order, we have
\begin{eqnarray}
\textrm{LO}:&H^0& \to b\bar{b}[^3S_1^{[8]}]+g, \nonumber \\
\textrm{NLO}:&H^0& \to b\bar{b}[^3S_1^{[8]}]+g~(\textrm{virtual}), \nonumber \\
&H^0& \to b\bar{b}[^3S_1^{[8]}]+g+g, \nonumber \\
&H^0& \to b\bar{b}[^3S_1^{[8]}]+u_g+\bar{u}_g~(\textrm{ghost}), \nonumber \\
&H^0& \to b\bar{b}[^3S_1^{[8]}]+q+\bar{q}, \nonumber \\
\textrm{NLO}^{*}:&H^0& \to b\bar{b}[^3S_1^{[8]}]+b+\bar{b}, \nonumber \\
&H^0& \to b\bar{b}[^3S_1^{[8]}]+c+\bar{c}. \label{3s18 channels}
\end{eqnarray}
The label ``$\textrm{NLO}^{*}$" denotes the heavy quark-antiquark pair associated processes, which are free of divergence. 
\item
In the cases of $n=^3S_1^{[1]},^1S_0^{[8]},^3P_J^{[8]}$, and $^3P_J^{[1]}$, the involved channels are
\begin{eqnarray}
&H^0& \to b\bar{b}[^3S_1^{[1]},^1S_0^{[8]},^3P_J^{[8]},^3P_J^{[1]}]+b+\bar{b}, \nonumber \\
&H^0& \to b\bar{b}[^1S_0^{[8]},^3P_J^{[8]},^3P_J^{[1]}]+g+g. \label{non-3s18 channels}
\end{eqnarray}
\end{itemize}
Typical Feynman diagrams corresponding to Eqs. (\ref{3s18 channels}) are presented in Figs. \ref{fig:Feyn1}, \ref{fig:Feyn2}, and \ref{fig:Feyn3}. The diagrams for $H^0 \to b\bar{b}[^3S_1^{[1]},^1S_0^{[8]},^3P_J^{[8]},^3P_J^{[1]}]+b+\bar{b}$ are the same with the first two diagrams of Fig. \ref{fig:Feyn3},  and the diagrams for $H^0 \to b\bar{b}[^1S_0^{[8]},^3P_J^{[8]},^3P_J^{[1]}]+g+g$ are the same with the ones in the first line of Fig. \ref{fig:Feyn2} excluding the 3-gluon vertex diagrams.

In the following, we will briefly present the formalisms for the NLO QCD corrections to $H^0 \to b\bar{b}[^3S_1^{[8]}]+g$ as well as the calculations for the tree-level process of $H^0 \to b\bar{b}[^3P_J^{[1]},^3P_J^{[8]}]+g+g$. The rest processes in Eq. (\ref{non-3s18 channels}) and the $\textrm{NLO}^{*}$ processes are both free of divergence, thus one can take the calculations directly according to the Feynman rules.

\subsection{NLO QCD corrections to $H^0 \to b\bar{b}[^3S_1^{[8]}]+g$}

To the next-to-leading order in $\alpha_s$, the SDC of the process of $H^0 \to b\bar{b}[^3S_1^{[8]}]+X_{\textrm{light-hadrons}}$ reads
\begin{eqnarray}
\hat{\Gamma}=\hat{\Gamma}_{\textrm{Born}}+\hat{\Gamma}_{\textrm{Virtual}}+\hat{\Gamma}_{\textrm{Real}}+\mathcal O(\alpha\alpha_s^3),
\end{eqnarray}
where
\begin{eqnarray}
&&\hat{\Gamma}_{\textrm{Virtual}}=\hat{\Gamma}_{\textrm{Loop}}+\hat{\Gamma}_{\textrm{CT}}, \nonumber \\
&&\hat{\Gamma}_{\textrm{Real}}=\hat{\Gamma}_{\textrm{S}}+\hat{\Gamma}_{\textrm{HC}}+\hat{\Gamma}_{\textrm{H}\overline{\textrm{C}}}.
\end{eqnarray}
$\hat{\Gamma}_{\textrm{Virtual}}$ is the virtual corrections, consisting of the contributions from the one-loop diagrams ($\hat{\Gamma}_{\textrm{Loop}}$) and the counterterms ($\hat{\Gamma}_{\textrm{CT}}$). $\hat{\Gamma}_{\textrm{Real}}$ stands for the real corrections, which includes the soft terms ($\hat{\Gamma}_{S}$), hard-collinear terms $(\hat{\Gamma}_{\textrm{HC}})$, and hard-noncollinear terms $(\hat{\Gamma}_{\textrm{H}\overline{\textrm{C}}})$. To isolate the ultraviolet (UV) and infrared (IR) divergences, we adopt the dimensional regularization with $D=4-2\epsilon$. The on-mass-shell (OS) scheme is employed to set the renormalization constants for the heavy quark mass ($Z_m$), heavy quark filed ($Z_2$), and gluon filed ($Z_3$). The modified minimal-subtraction ($\overline{MS}$) scheme is used for the QCD gauge coupling ($Z_g$). The renormalization constants are \cite{Gong:2007db},
\begin{eqnarray}
\delta Z_{m}^{OS}&=& -3 C_{F} \frac{\alpha_s N_{\epsilon}}{4\pi}\left[\frac{1}{\epsilon_{\textrm{UV}}}-\gamma_{E}+\textrm{ln}\frac{4 \pi \mu_r^2}{m_b^2}+\frac{4}{3}\right], \nonumber \\
\delta Z_{2}^{OS}&=& - C_{F} \frac{\alpha_s N_{\epsilon}}{4\pi}\left[\frac{1}{\epsilon_{\textrm{UV}}}+\frac{2}{\epsilon_{\textrm{IR}}}-3 \gamma_{E}+3 \textrm{ln}\frac{4 \pi \mu_r^2}{m_b^2}+4\right], \nonumber \\
\delta Z_{3}^{\overline{MS}}&=& \frac{\alpha_s N_{\epsilon}}{4\pi}\left[(\beta_{0}^{'}-2 C_{A})(\frac{1}{\epsilon_{\textrm{UV}}}-\frac{1}{\epsilon_{\textrm{IR}}})-\frac{4}{3}T_F(\frac{1}{\epsilon_{\textrm{UV}}}-\gamma_E+\textrm{ln}\frac{4\pi\mu_r^2}{m_c^2}) \right. \nonumber\\
&& \left. -\frac{4}{3}T_F(\frac{1}{\epsilon_{\textrm{UV}}}-\gamma_E+\textrm{ln}\frac{4\pi\mu_r^2}{m_b^2})\right], \nonumber \\
\delta Z_{g}^{\overline{MS}}&=& -\frac{\beta_{0}}{2}\frac{\alpha_s N_{\epsilon}}{4\pi}\left[\frac{1} {\epsilon_{\textrm{UV}}}-\gamma_{E}+\textrm{ln}(4\pi)\right], \label{CT}
\end{eqnarray}
where $\gamma_E$ is the Euler's constant, $N_{\epsilon}= \Gamma[1-\epsilon] /({4\pi\mu_r^2}/{(4m_b^2)})^{\epsilon}$, $\beta_{0}=\frac{11}{3}C_A-\frac{4}{3}T_Fn_f$ is the one-loop coefficient of the $\beta$-function, and $\beta_{0}^{'}=\frac{11}{3}C_A-\frac{4}{3}T_Fn_{lf}$. $n_f$ and $n_{lf}$ are the number of active quark flavors and light quark flavors, respectively. In ${\rm SU}(3)$, the color factors are given by $T_F=\frac{1}{2}$, $C_F=\frac{4}{3}$, and $C_A=3$. The two-cutoff slicing strategy is utilized to subtract the IR divergences in $\Gamma_{\textrm{Real}}$ \cite{Harris:2001sx}.

\subsection{$H^0 \to b\bar{b}[^3P_J^{[1]},^3P_J^{[8]}]+g+g$}

Taking $^3P_J^{[1]}$ as an example \footnote{Since the process of $H^0 \to b\bar{b}[^3S_1^{[1]}]+g$ is forbidden, for the $H^0 \to b\bar{b}[^3P_J^{[8]}]+g+g$ case, the calculation formalism is almost the same except the color factor.}, we first divide $\Gamma_{H^0 \to b\bar{b}[^3P_J^{[1]}]+g+g}$ into two ingredients,
\begin{eqnarray}
&&d\Gamma_{H^0 \to b\bar{b}[^3P_J^{[1]}]+g+g}=d\hat{\Gamma}_{^3P_J^{[1]}} \langle \mathcal O^{\chi_b}(^3P_J^{[1]}) \rangle+d\hat{\Gamma}_{^3S_1^{[8]}}^{LO} \langle \mathcal O^{\chi_b}(^3S_1^{[8]}) \rangle ^{NLO},
\end{eqnarray}
then one can obtain
\begin{eqnarray}
d\hat{\Gamma}_{^3P_J^{[1]}} \langle \mathcal O^{\chi_b}(^3P_J^{[1]}) \rangle&=&d\Gamma_{H^0 \to b\bar{b}[^3P_J^{[1]}]+g+g} -d\hat{\Gamma}_{^3S_1^{[8]}}^{LO} \langle \mathcal O^{\chi_b}(^3S_1^{[8]}) \rangle ^{NLO} \nonumber \\
&=&d{\Gamma}_F+d{\Gamma}_S-d\hat{\Gamma}_{^3S_1^{[8]}}^{LO} \langle \mathcal O^{\chi_b}(^3S_1^{[8]}) \rangle ^{NLO}. \label{3pj1 SDC}
\end{eqnarray}
$d\Gamma_F(=d\hat{\Gamma}_F \langle \mathcal O^{\chi_b}(^3P_J^{[1]}) \rangle)$ is the finite term in $d\Gamma_{H^0 \to b\bar{b}[^3P_J^{[1]}]+g+g}$ and $d\Gamma_S$ is the soft part which can be written as
\begin{eqnarray}
&&d{\Gamma}_S=-\frac{\alpha_s}{3 \pi m_b^2} u^{s}_\epsilon \frac{N_c^2-1}{N_c^2} d\hat{\Gamma}^{LO}_{^3S_1^{[8]}} \langle \mathcal O^{\chi_b}(^3P_J^{[1]}) \rangle, \label{3pj1 soft}
\end{eqnarray}
where
\begin{eqnarray}
u^{s}_\epsilon=\frac{1}{\epsilon_{IR}}+\frac{E}{|\textbf{p}|} \textrm{ln}(\frac{E+|\textbf{p}|}{E-|\textbf{p}|}) + \textrm{ln}(\frac{4 \pi \mu_r^2}{s\delta_s^2})-\gamma_E-\frac{1}{3}. \label{us}
\end{eqnarray}
$N_c$ is identical to 3 for $\textrm{SU}(3)$ gauge field. $E$ and $\textbf{p}$ denote the energy and 3-momentum of $\chi_b$, respectively. $\delta_s$ is the usual ``soft cut" employed to impose an amputation on the energy of the emitted gluon. Regarding $\langle \mathcal O^{\chi_b}(^3S_1^{[8]}) \rangle ^{NLO}$, under the dimensional regularization scheme as is adopted in \cite{Jia:2014jfa}, we have
\begin{eqnarray}
\langle \mathcal O^{\chi_b}(^3S_1^{[8]}) \rangle ^{NLO}=-\frac{\alpha_s}{3 \pi m_b^2} u^{c}_\epsilon \frac{N_c^2-1}{N_c^2} \langle \mathcal O^{\chi_b}(^3P_J^{[1]}) \rangle. \label{3s18to3pj1}\
\end{eqnarray}
Then the third term in Eq. (\ref{3pj1 SDC}) can be written as
\begin{eqnarray}
d\hat{\Gamma}_{^3S_1^{[8]}}^{LO} \langle \mathcal O^{\chi_b}(^3S_1^{[8]}) \rangle ^{NLO})=-\frac{\alpha_s}{3 \pi m_b^2} u^{c}_\epsilon \frac{N_c^2-1}{N_c^2} d\hat{\Gamma}^{LO}_{^3S_1^{[8]}} \langle \mathcal O^{\chi_b}(^3P_J^{[1]}) \rangle, \label{3s18to3pj1II}
\end{eqnarray}
where, on the basis of $\mu_{\Lambda}$-cutoff scheme \cite{Jia:2014jfa}, $u^{c}_\epsilon$ has the following form
\begin{eqnarray}
u^{c}_\epsilon=\frac{1}{\epsilon_{IR}}-\gamma_E-\frac{1}{3}+ \textrm{ln}(\frac{4 \pi \mu_r^2}{\mu_{\Lambda}^2}). \label{uc}
\end{eqnarray}
$\mu_{\Lambda}$ is the upper bound of the integrated gluon energy, rising from the renormalization of the LDME. Substituting Eqs. (\ref{3pj1 soft}) and (\ref{3s18to3pj1II}) into Eq. (\ref{3pj1 SDC}), the soft singularities in $d{\Gamma}_S$ and $d\hat{\Gamma}_{^3S_1^{[8]}}^{LO} \langle \mathcal O^{\chi_c}(^3S_1^{[8]}) \rangle ^{NLO}$ cancel each other, consequently leading to
\begin{eqnarray}
d\hat{\Gamma}_{^3P_J^{[1]}} \langle \mathcal O^{\chi_b}(^3P_J^{[1]}) \rangle
&=&\left[d\hat{{\Gamma}}_F+\frac{\alpha_s}{3 \pi m_b^2} (u^{c}_\epsilon-u^{s}_\epsilon) \frac{N_c^2-1}{N_c^2} d\hat{\Gamma}^{LO}_{^3S_1^{[8]}}\right]\langle \mathcal O^{\chi_b}(^3P_J^{[1]}) \rangle \\ \nonumber
&=&(d\hat{{\Gamma}}_F+d\hat{{\Gamma}}^{*})\langle \mathcal O^{\chi_b}(^3P_J^{[1]}) \rangle .
\end{eqnarray}

\begin{figure*}
\includegraphics[width=0.49\textwidth]{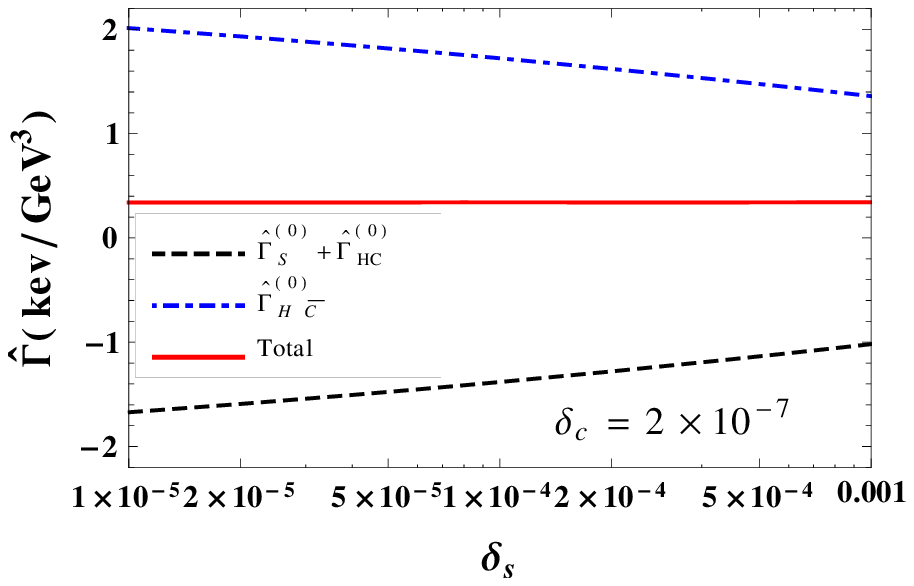}
\includegraphics[width=0.49\textwidth]{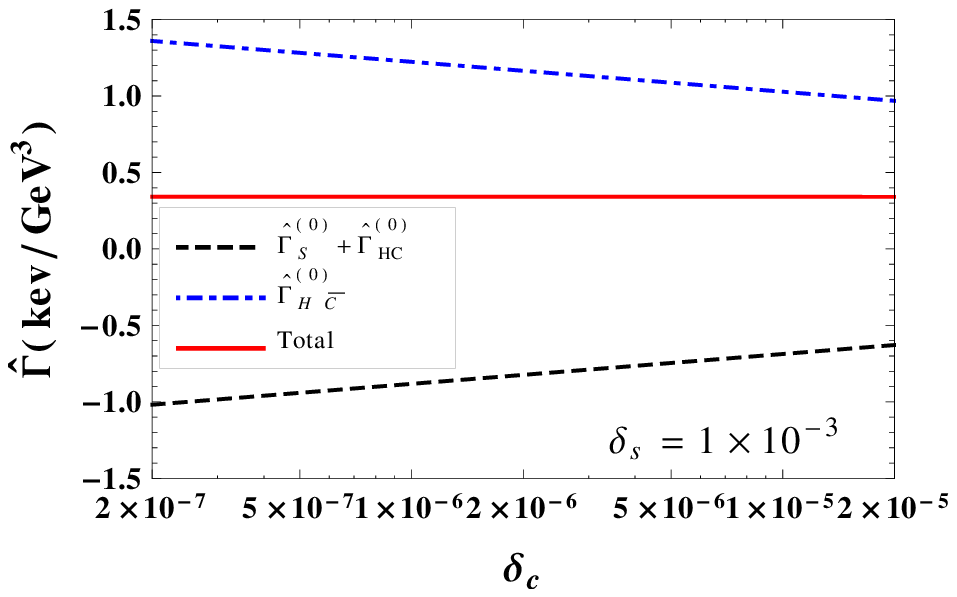}
\includegraphics[width=0.49\textwidth]{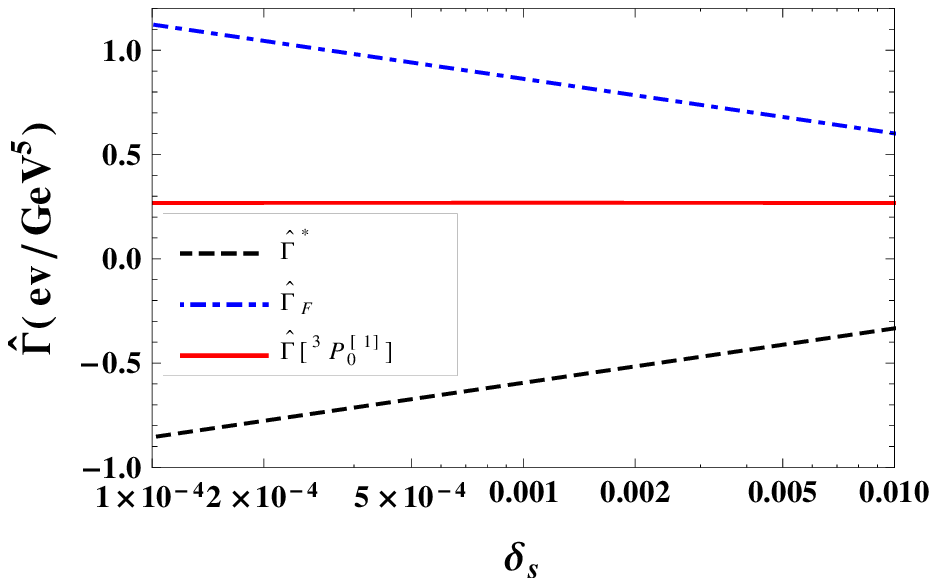}
\includegraphics[width=0.49\textwidth]{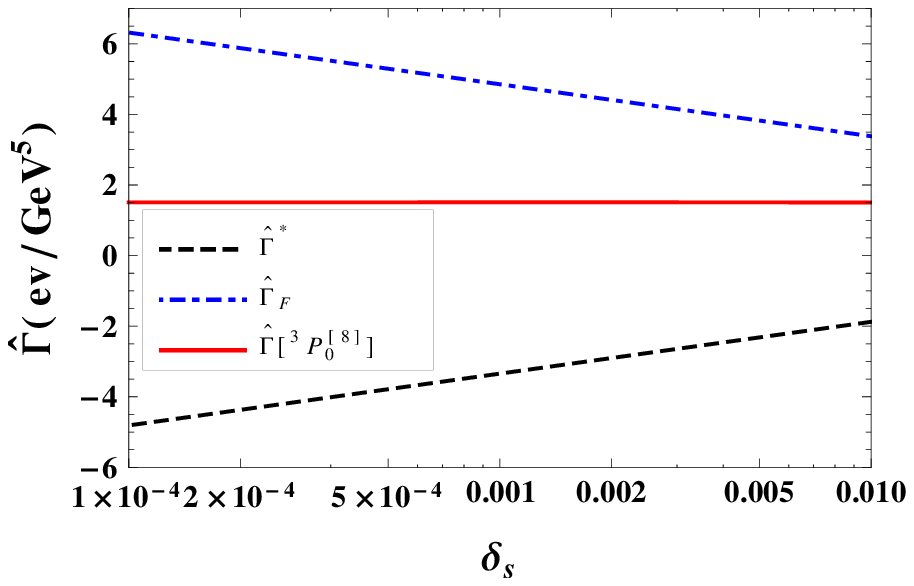}
\caption{\label{fig:cut}
The verification of the independence on the cutoff parameters of $\delta_{s,c}$ for the SDCs of $^3S_1^{[8]}$ (the upper two diagrams) and $^3P_0^{[1,8]}$ (the lower two diagrams), respectively. The superscript ``(0)" denotes the $\epsilon^{0}$-order terms.}
\end{figure*}

The package $\textrm{M}\scriptsize{\textrm{ALT@FDC}}$ that has been adopted in several heavy quarkonium related processes \cite{Sun:2018hpb,Gong:2016jiq,Feng:2017bdu,Zhang:2017dia,Sun:2017nly,Sun:2017wxk,Jiang:2018wmv} is used to deal with $\hat{\Gamma}_{\textrm{Virtual}}$, $\hat{\Gamma}_{\textrm{\textrm{S}}}$, and $\hat{\Gamma}_{\textrm{\textrm{HC}}}$. To calculate the hard-noncollinear part of the real corrections, $\hat{\Gamma}_{\textrm{H}\overline{\textrm{C}}}$, we employ the $\textrm{FDC}$~\cite{Wang:2004du} package. Both the cancellation of the $\epsilon^{-2(-1)}$-order divergence and the independence on cutoff ($\delta_{s,c}$) have been checked carefully. Taking $^3S_1^{[8]}$ and $^3P_J^{[1,8]}$ $(J=0)$ as an example, the verification of the independence on the cutoff parameters of $\delta_{s,c}$ is shown in Fig. \ref{fig:cut}. The $J=1,2$ cases are not presented here since they are quite similar to the $J=0$ case.

\section{Phenomenological results}

For the numerical calculations, we take $\alpha=1/128$, $m_c=1.5$ GeV, $m_b=4.9$ GeV, $m_t=173$ GeV, $m_{W}=80.4$ GeV, and $m_{H^0}=125$ GeV. The light quarks $q$ and $\bar{q}$ $(q=u,d,s)$ are regarded as massless. For the NLO corrections to $H^0 \to b\bar{b}[^3S_1^{[8]}]+g$ and calculating other $\alpha_s^2-$order processes, we employ the two-loop $\alpha_s$ running. The one-loop $\alpha_s$ running is adopted for the LO cases. The mixed feeddown scheme of $\chi_{bJ}(3P) \to \Upsilon(nS)$ in Ref. \cite{Han:2014kxa} is used and the value of $\mu_{\Lambda}$ is taken as $m_b$, thus the CO LDMEs in Table 4 of Ref. \cite{Feng:2015wka} are chosen to achieve the numerical results. For the CS cases, $^3S_1^{[1]}$ and $^3P_J^{[1]}$, the LDMEs are related to the radial wave functions at the origin ($n,m=1,2,3$):
\begin{eqnarray}
\frac{\langle \mathcal O^{\Upsilon(nS)}(^3S_1^{[1]}) \rangle}{6N_c}&=&\frac{1}{4\pi}|R_{\Upsilon(nS)}(0)|^2, \\ \nonumber
\frac{\langle \mathcal O^{\chi_{bJ}(mP)}(^3P_J^{[1]}) \rangle}{2N_c}&=&(2J+1)\frac{3}{4\pi}|R^{'}_{\chi_b(mP)}(0)|^2,
\end{eqnarray}
where $|R_{\Upsilon(nS)}(0)|^2$ and $|R^{'}_{\chi_b(mP)}(0)|^2$ are taken as \cite{Eichten:1995ch}
\begin{eqnarray}
&&|R_{\Upsilon(1S)}(0)|^2=6.477~\textrm{GeV}^3,~~~|R_{\Upsilon(2S)}(0)|^2=3.234~\textrm{GeV}^3, \\ \nonumber
&&|R_{\Upsilon(3S)}(0)|^2=2.474~\textrm{GeV}^3, \\ \nonumber
&&|R^{'}_{\chi_b(1P)}(0)|^2=1.417~\textrm{GeV}^5,~~~|R^{'}_{\chi_b(2P)}(0)|^2=1.653~\textrm{GeV}^5, \\ \nonumber
&&|R^{'}_{\chi_b(3P)}(0)|^2=1.794~\textrm{GeV}^5. \nonumber
\end{eqnarray}
Branching ratios of $\chi_{bJ}(mP) \to \Upsilon(nS)$, $\Upsilon(nS) \to \chi_{bJ}(mP)$, $\Upsilon(3S) \to \Upsilon(2S)$, $\Upsilon(3S) \to \Upsilon(1S)$, and $\Upsilon(2S) \to \Upsilon(1S)$ can be found in Refs. \cite{Gong:2013qka,Han:2014kxa,Feng:2015wka}.
\begin{table*}[htb]
\caption{The SDC of $^3S_1^{[8]}$ (in units of $\textrm{kev}/\textrm{GeV}^{3}$).}
\label{SDC3s18}
\begin{tabular}{ccccccccc}
\hline\hline
$\mu_r$ & $\textrm{LO}$ & $\textrm{NLO}$ & $\textrm{NLO}^{*}_{b\bar{b}}$ & $\textrm{NLO}^{*}_{c\bar{c}}$ & $\textrm{Total}$\\ \hline
$2m_b$ & $8.79 \times 10^{-2}$ & $0.340$ & $0.568$ & $6.44 \times 10^{-2}$ & $0.97$\\
$m_{H^0}$ & $5.42 \times 10^{-2}$ & $0.170$ & $0.223$ & $2.53 \times 10^{-2}$ & $0.42$\\ \hline\hline
\end{tabular}
\end{table*}

Before presenting the phenomenological results, we first take a look at the effect of the QCD corrections to the process of $H^0 \to b\bar{b}[^3S_1^{[8]}]+g$, presented in Table \ref{SDC3s18}. We see that the newly calculated NLO terms increase the LO results by about 3-4 times, causing a $40\%$ enhancement on the total $^3S_1^{[8]}$ contributions ($\textrm{LO}+\textrm{NLO}^{*}_{b\bar{b},c\bar{c}}$). This is consistent with the lesson we learn from $Z^0$ decay \cite{Sun:2018hpb}. The $^3S_1^{[8]}$ state may provide significant (even dominant) contributions to $\Gamma_{H^0 \to \Upsilon,\chi_b+X}$, thus the newly introduced NLO ingredient is of great essence in achieving the phenomenological conclusions.

\subsection{$\chi_b(3P,2P,1P)$}

\begin{table*}[htb]
\caption{The decay widths of $H^0 \to \chi_{bJ}(3P)+X$ (in units of ev). The superscripts ``DR" and ``FD" denote the direct production processes and feeddown effects, respectively.}
\label{xb3p}
\begin{tabular}{cccccccccccc}
\hline\hline
$\chi_{bJ}$ & $\mu_r$ & $^3S_1^{[8]}$ & $^3P_0^{[1]}|_{gg}$ & $^3P_0^{[1]}|_{b\bar{b}}$ & $\Gamma_{\textrm{DR}}$ & $\Gamma_{\textrm{FD}}^{\Upsilon}$ & $\Gamma_{\textrm{Total}}$ & $\textrm{Br}(\times 10^{-6})$\\ \hline
$J=0$ & $2m_b$ & $6.22$ & $0.69$ & $12.9$ & $19.8$ & $-$ & $19.8$ & $4.71$\\
$~$ & $m_{H^0}$ & $2.67$ & $0.27$ & $5.06$ & $8.00$ & $-$ & $8.00$ & $1.90$\\ \hline
$J=1$ & $2m_b$ & $18.7$ & $0.91$ & $14.0$ & $33.6$ & $-$ & $33.6$ & $8.00$\\
$~$ & $m_{H^0}$ & $8.03$ & $0.36$ & $5.49$ & $13.9$ & $-$ & $13.9$ & $3.31$\\ \hline
$J=2$ & $2m_b$ & $31.1$ & $4.09$ & $5.06$ & $40.3$ & $-$ & $40.3$ & $9.60$\\
$~$ & $m_{H^0}$ & $13.4$ & $1.60$ & $1.99$ & $17.0$ & $-$ & $17.0$ & $4.05$\\ \hline
\end{tabular}
\end{table*}

\begin{table*}[htb]
\caption{The decay widths of $H^0 \to \chi_{bJ}(2P)+X$ (in units of ev). The superscripts ``DR" and ``FD" denote the direct production processes and feeddown effects, respectively.}
\label{xb2p}
\begin{tabular}{cccccccccccc}
\hline\hline
$\chi_{bJ}$ & $\mu_r$ & $^3S_1^{[8]}$ & $^3P_0^{[1]}|_{gg}$ & $^3P_0^{[1]}|_{b\bar{b}}$ & $\Gamma_{\textrm{DR}}$ & $\Gamma_{\textrm{FD}}^{\Upsilon(3S)}$ & $\Gamma_{\textrm{Total}}$ & $\textrm{Br}(\times 10^{-6})$\\ \hline
$J=0$ & $2m_b$ & $4.86$ & $0.64$ & $11.9$ & $17.4$ & $5.83$ & $23.2$ & $5.52$\\
$~$ & $m_{H^0}$ & $2.09$ & $0.25$ & $4.66$ & $7.00$ & $2.33$ & $9.33$ & $2.22$\\ \hline
$J=1$ & $2m_b$ & $14.6$ & $0.84$ & $12.9$ & $28.3$ & $12.5$ & $40.8$ & $9.71$\\
$~$ & $m_{H^0}$ & $6.27$ & $0.33$ & $5.06$ & $11.7$ & $4.98$ & $16.7$ & $3.98$\\ \hline
$J=2$ & $2m_b$ & $24.3$ & $3.76$ & $4.66$ & $32.7$ & $12.9$ & $45.6$ & $10.9$\\
$~$ & $m_{H^0}$ & $10.5$ & $1.48$ & $1.83$ & $13.8$ & $5.18$ & $19.0$ & $4.52$\\ \hline
\end{tabular}
\end{table*}

\begin{table*}[htb]
\caption{The decay widths of $H^0 \to \chi_{bJ}(1P)+X$ (in units of ev). The superscripts ``DR" and ``FD" denote the direct production processes and feeddown effects, respectively.}
\label{xb1p}
\begin{tabular}{cccccccccccc}
\hline\hline
$\chi_{bJ}$ & $\mu_r$ & $^3S_1^{[8]}$ & $^3P_0^{[1]}|_{gg}$ & $^3P_0^{[1]}|_{b\bar{b}}$ & $\Gamma_{\textrm{DR}}$ & $\Gamma_{\textrm{FD}}^{\Upsilon(2S)}$ & $\Gamma_{\textrm{Total}}$ & $\textrm{Br}(\times 10^{-6})$\\ \hline
$J=0$ & $2m_b$ & $3.76$ & $0.54$ & $10.2$ & $14.5$ & $5.94$ & $20.4$ & $4.86$\\
$~$ & $m_{H^0}$ & $1.62$ & $0.21$ & $4.00$ & $5.83$ & $2.39$ & $8.22$ & $1.96$\\ \hline
$J=1$ & $2m_b$ & $11.3$ & $0.72$ & $11.1$ & $23.1$ & $10.8$ & $33.9$ & $8.07$\\
$~$ & $m_{H^0}$ & $4.85$ & $0.28$ & $4.34$ & $9.47$ & $4.33$ & $13.8$ & $3.29$\\ \hline
$J=2$ & $2m_b$ & $18.8$ & $3.23$ & $4.00$ & $26.0$ & $11.2$ & $37.2$ & $8.86$\\
$~$ & $m_{H^0}$ & $8.08$ & $1.27$ & $1.57$ & $10.9$ & $4.49$ & $15.4$ & $3.67$\\ \hline
\end{tabular}
\end{table*}

The NRQCD predictions on the decay width of $H^0 \to \chi_{bJ}(3P,2P,1P)+X$ are listed in Tables. \ref{xb3p}, \ref{xb2p}, and \ref{xb1p}. In order to show the relative importance of different production channels in a wide range of $\mu_r$, we provide the predictions at $\mu_r=2m_b$ and $\mu_r=m_{H^0}$ simultaneously. It is noticed that the branching ratios for $H^0 \to \chi_{bJ}(3P,2P,1P)+X$ are calculated to be on the order of $10^{-6}-10^{-5}$, indicating the probability of these processes to be observed at the HE-LHC,  HL-LHC, and other colliders in near future. In addition to the direct production processes that are dominant, the feeddown effects via the higher excited states, e.g., $\Upsilon(2S)$ and $\Upsilon(1S)$, are also significant, accounting for about $30\%$ of the total decay width, as is shown in Table \ref{xb2p} and Table \ref{xb1p}. The direct productions consist of two parts, i.e. the CS state $^3P_J^{[1]}$ and the CO state $^3S_1^{[8]}$. 
\begin{itemize}
\item
For the CS cases, the processes of $H^{0} \to b\bar{b}[^3P_J^{[1]}]+b+\bar{b}$ ($``b\bar{b}"$) serve as the leading role in the total CS prediction due to the $b$-quark fragmentation mechanism. However, the light hadrons associated process $H^{0} \to b\bar{b}[^3P_J^{[1]}]+g+g$ ($``gg"$) can also provide non-negligible contributions. To be specific, for $^3P_0^{[1]}$ and $^3P_1^{[1]}$ states, the contribution of the $``gg"$ channel enhance the $``b\bar{b}"$ cases by about $5\%$ and $7\%$, respectively. Moreover, for the $^3P_2^{[1]}$ case, the $``gg"$ contribution can surprisingly reach up to about $81\%$ of the $``b\bar{b}"$ contribution. Therefore, to achieve a sound estimate, besides $H^{0} \to b\bar{b}[^3P_J^{[1]}]+b+\bar{b}$, the contributions of $H^{0} \to b\bar{b}[^3P_J^{[1]}]+g+g$ must be also taken into consideration.
\item
Regarding the CO cases, including the $^3S_1^{[8]}$ state contributions can significantly enlarge the predicted decay width. Taking $\chi_{bJ}(3P)$ as an example, when $\mu_r=2m_b$, the $^3S_1^{[8]}$ contributions account for about $31\%,56\%$, and $77\%$ of $\Gamma_{\textrm{DR}}$, corresponding to $J=0,1$, and $2$, respectively. As for the $\chi_b(2P)$ and $\chi_b(1P)$ cases, the proportions are about $28\%, 52\%, 74\%$ and $26\%, 49\%, 72\%$, respectively. 
\end{itemize}

\begin{table*}[htb]
\caption{The ratios of $\Gamma_{\chi_{b2}} / \Gamma_{\chi_{b0}}$ and $\Gamma_{\chi_{b2}} / \Gamma_{\chi_{b1}}$. ``CS" denotes the sum of the CS direct ($^3P_J^{[1]}$) and feeddown ($^3S_1^{[1]}$) contributions , while ``NR" means the NRQCD results including both CS and CO contributions. $\mu_r$ is varied in $\left[2m_b,m_{H^{0}}\right]$.}
\label{rchib}
\begin{tabular}{ccccccccc}
\hline\hline
$~$ & $\frac{\Gamma_{\chi_{b2}}}{\Gamma_{\chi_{b0}}}|_{3P}$ & $\frac{\Gamma_{\chi_{b2}}}{\Gamma_{\chi_{b1}}}|_{3P}$ & $\frac{\Gamma_{\chi_{b2}}}{\Gamma_{\chi_{b0}}}|_{2P}$ & $\frac{\Gamma_{\chi_{b2}}}{\Gamma_{\chi_{b1}}}|_{2P}$ & $\frac{\Gamma_{\chi_{b2}}}{\Gamma_{\chi_{b0}}}|_{1P}$ & $\frac{\Gamma_{\chi_{b2}}}{\Gamma_{\chi_{b1}}}|_{1P}$\\ \hline
$\textrm{CS}$ & $0.674$ & $0.613$ & $1.097$ & $0.794$ & $1.029$ & $0.786$\\
$\textrm{NR}$ & $2.035 \sim 2.125$ & $1.199 \sim 1.223$ & $1.966 \sim 2.036$ & $1.118 \sim 1.138$ & $1.824 \sim 1.873$ & $1.097 \sim 1.116$\\ \hline\hline
\end{tabular}
\end{table*}

In addition to the large contributions to the total decay width, the $^3S_1^{[8]}$ state also has crucial effect on the ratios of $\Gamma_{\chi_{b2}} / \Gamma_{\chi_{b0}}$ and $\Gamma_{\chi_{b2}} / \Gamma_{\chi_{b1}}$, as shown in Table \ref{rchib}, where the feeddown effects have been incorporated. Since the dependence of the CS channels, $``gg"$ and $``b\bar{b}"$, on $\mu_r$ is only in the strong coupling constants $\alpha_s$, varying $\mu_r$ of course does not affect the ratios. However, for the CO cases, due to the NLO corrections to $H^0 \to b\bar{b}[^3S_1^{[8]}]+g$, the form of the dependence on $\mu_r$ is not only $\alpha_s$. Although varying $\mu_r$ in $\left[2m_b,m_{H^{0}}\right]$ greatly influence the total decay widths, the ratios of $\Gamma_{\chi_{b2}} / \Gamma_{\chi_{b0}}$ and $\Gamma_{\chi_{b2}} / \Gamma_{\chi_{b1}}$ are quite insensitive to the choice of $\mu_r$. Taking $\frac{\Gamma_{\chi_{b2}}}{\Gamma_{\chi_{b0}}}|_{3P}$ for example, when $\mu_r$ is varied from $2m_b$ (9.8 GeV) to $m_{H^{0}}$ (125 GeV), the ratios just increase by about $4\%$. In addition, the differences between the CS and NRQCD results are rather conspicuous, which can be regarded as an outstanding probe to distinguish between the two heavy quarkonium production mechanism.

\subsection{$\Upsilon(3S,2S,1S)$}

\begin{table*}[htb]
\caption{The decay widths of $H^0 \to \Upsilon(3S)+X$ (in units of ev). The superscripts ``DR" and ``FD" denote the direct production processes and feeddown effects, respectively.}
\label{upsilon3s}
\begin{tabular}{cccccccccccc}
\hline\hline
$\mu_r$ & $^3S_1^{[8]}$ & $^1S_0^{[8]}$ & $^3P_J^{[8]}$ & $^3S_1^{[1]}$ & $\Gamma_{\textrm{DR}}$ & $\Gamma_{\textrm{FD}}^{\chi_b(3P)}$ & $\Gamma_{\textrm{FD}}^{\Upsilon}$ & $\Gamma_{\textrm{Total}}$ & $\textrm{Br}(\times 10^{-5})$\\ \hline
$2m_b$ & $14.8$ & $-3.29 \times 10^{-2}$ & $-1.69 \times 10^{-2}$ & $77.1$ & $91.9$ & $6.96$ & $-$ & $98.9$ & $2.35$\\
$m_{H^0}$ & $6.35$ & $-1.29 \times 10^{-2}$ & $-6.62 \times 10^{-3}$  & $30.3$ & $36.6$ & $2.89$ & $-$ & $39.5$ & $0.94$\\ \hline\hline
\end{tabular}
\end{table*}

\begin{table*}[htb]
\caption{The decay widths of $H^0 \to \Upsilon(2S)+X$ (in units of ev). The superscripts ``DR" and ``FD" denote the direct production processes and feeddown effects, respectively.}
\label{upsilon2s}
\begin{tabular}{cccccccccccc}
\hline\hline
$\mu_r$ & $^3S_1^{[8]}$ & $^1S_0^{[8]}$ & $^3P_J^{[8]}$ & $^3S_1^{[1]}$ & $\Gamma_{\textrm{DR}}$ & $\Gamma_{\textrm{FD}}^{\chi_b(2,3P)}$ & $\Gamma_{\textrm{FD}}^{\Upsilon(3S)}$ & $\Gamma_{\textrm{Total}}$ & $\textrm{Br}(\times 10^{-5})$\\ \hline
$2m_b$ & $28.6$ & $-0.11$ & $0.47$ & $101$ & $130$ & $16.1$ & $10.5$ & $157$ & $3.74$\\
$m_{H^0}$ & $12.3$ & $-4.23 \times 10^{-2}$ & $0.19$  & $39.6$ & $52.1$ & $6.60$ & $4.19$ & $62.9$ & $1.50$\\ \hline\hline
\end{tabular}
\end{table*}

\begin{table*}[htb]
\caption{The decay widths of $H^0 \to \Upsilon(1S)+X$ (in units of ev). The superscripts ``DR" and ``FD" denote the direct production processes and feeddown effects, respectively.}
\label{upsilon1s}
\begin{tabular}{cccccccccccc}
\hline\hline
$\mu_r$ & $^3S_1^{[8]}$ & $^1S_0^{[8]}$ & $^3P_J^{[8]}$ & $^3S_1^{[1]}$ & $\Gamma_{\textrm{DR}}$ & $\Gamma_{\textrm{FD}}^{\chi_b(1,2,3P)}$ & $\Gamma_{\textrm{FD}}^{\Upsilon(2,3S)}$ & $\Gamma_{\textrm{Total}}$ & $\textrm{Br}(\times 10^{-5})$\\ \hline
$2m_b$ & $4.57$ & $2.12$ & $-0.83$ & $202$ & $208$ & $27.3$ & $48.0$ & $283$ & $6.74$\\
$m_{H^0}$ & $1.96$ & $0.83$ & $-0.32$  & $79.3$ & $81.8$ & $12.1$ & $19.3$ & $113$ & $2.69$\\ \hline\hline
\end{tabular}
\end{table*}

The NRQCD predictions on the decay width of $H^0 \to \Upsilon(3S,2S,1S)+X$ are presented in Tables. \ref{upsilon3s}, \ref{upsilon2s}, and \ref{upsilon1s}, respectively. In these tables, one would see that the branching ratios of the inclusive productions of $\Upsilon(3S,2S,1S)$ via $H^0$ decay are about $10^{-5}-10^{-4}$, indicating the potential to be detected at the high energy collider. For $H^0 \to \Upsilon(3S,2S,1S)+X$, the feeddown contributions from the higher excited states are remarkable, accounting for about $7\%,17\%$, and $27\%$ of the total decay widths of $\Upsilon(3S)$, $\Upsilon(2S)$, and $\Upsilon(1S)$, respectively. Regarding the direct productions, the main contributions come from the CS state, $^3S_1^{[1]}$, via the heavy-quark pair associated process. The CO states can also provide considerable contributions, which account for about $16\%,22\%$, and $3\%$ on $\Gamma_{\textrm{DR}}$ of $\Upsilon(3S)$, $\Upsilon(2S)$, and $\Upsilon(1S)$, respectively.

In addition to the total decay width, we also calculate the ratios of $\Gamma_{\Upsilon(2S)}/\Gamma_{\Upsilon(3S)}$ and $\Gamma_{\Upsilon(1S)}/\Gamma_{\Upsilon(3S)}$. By varying $\mu_r$ in $\left[2m_b,m_{H^{0}}\right]$, we have
\begin{eqnarray}
\textrm{CS}:&&\Gamma_{\Upsilon(2S)}/\Gamma_{\Upsilon(3S)}=1.471, \nonumber \\
&&\Gamma_{\Upsilon(1S)}/\Gamma_{\Upsilon(3S)}=3.170, \nonumber \\
\textrm{NR}:&&\Gamma_{\Upsilon(2S)}/\Gamma_{\Upsilon(3S)}=1.587 \sim 1.592, \nonumber \\
&&\Gamma_{\Upsilon(1S)}/\Gamma_{\Upsilon(3S)}=2.861 \sim 2.862,
\end{eqnarray}
where ``CS" denotes the sum of the CS direct ($^3P_J^{[1]}$) and feeddown ($^3S_1^{[1]}$) contributions , while ``NR" is the total results including both CS and CO contributions. The difference between the CS and NRQCD predictions reflects that the CO influence on $\Gamma_{\Upsilon(2S)}/\Gamma_{\Upsilon(3S)}$ and $\Gamma_{\Upsilon(1S)}/\Gamma_{\Upsilon(3S)}$ is moderate.

Finally, to serve as a useful reference, we analyze the uncertainties of the predictions due to the choices of the renormalization scale $\mu_r$, Higgs mass $m_{H^{0}}$, the bottom quark mass $m_b$, and the CO LDMEs. 
\begin{itemize}
\item
For $\mathcal B_{\chi_b(3P,2P,1P)} (\times 10^{-6})$
\begin{eqnarray}
&&\mathcal B_{H^0 \to \chi_{b0}(3P)+X}=2.34^{+0.61+0.03+0.25+0.14}_{-0.43-0.03-0.22-0.14}, \nonumber \\
&&\mathcal B_{H^0 \to \chi_{b1}(3P)+X}=4.03^{+1.02+0.06+0.39+0.41}_{-0.69-0.06-0.33-0.41}, \nonumber \\
&&\mathcal B_{H^0 \to \chi_{b2}(3P)+X}=4.90^{+1.21+0.07+0.37+0.68}_{-0.86-0.07-0.32-0.68}, \nonumber \\
&&\mathcal B_{H^0 \to \chi_{b0}(2P)+X}=2.73^{+0.71+0.04+0.27+0.12}_{-0.51-0.04-0.23-0.12}, \nonumber \\
&&\mathcal B_{H^0 \to \chi_{b1}(2P)+X}=4.85^{+1.25+0.07+0.42+0.33}_{-0.89-0.07-0.37-0.33}, \nonumber \\
&&\mathcal B_{H^0 \to \chi_{b2}(2P)+X}=5.50^{+1.38+0.08+0.39+0.50}_{-0.99-0.08-0.34-0.50}, \nonumber \\
&&\mathcal B_{H^0 \to \chi_{b0}(1P)+X}=2.40^{+0.63+0.03+0.23+0.06}_{-0.45-0.03-0.20-0.06}, \nonumber \\
&&\mathcal B_{H^0 \to \chi_{b1}(1P)+X}=4.02^{+1.03+0.06+0.35+0.14}_{-0.74-0.06-0.31-0.14}, \nonumber \\
&&\mathcal B_{H^0 \to \chi_{b2}(1P)+X}=4.47^{+1.13+0.06+0.32+0.20}_{-0.80-0.06-0.28-0.20}. \label{uncertaity chib} 
\end{eqnarray}
\item
And for $\mathcal B_{\Upsilon(3S,2S,1S)} (\times 10^{-5})$
\begin{eqnarray}
&&\mathcal B_{H^0 \to \Upsilon(3S)+X}=1.16^{+0.31+0.02+0.07+0.06}_{-0.22-0.02-0.06-0.06}, \nonumber \\
&&\mathcal B_{H^0 \to \Upsilon(2S)+X}=1.84^{+0.48+0.03+0.11+0.09}_{-0.34-0.03-0.10-0.09}, \nonumber \\
&&\mathcal B_{H^0 \to \Upsilon(1S)+X}=3.32^{+0.89+0.05+0.20+0.11}_{-0.63-0.05-0.18-0.11}, \label{uncertaity upsilon}
\end{eqnarray} 
\end{itemize} 
where the four columns are the uncertainties caused by $\mu_r$, $m_{H^{0}}$, $m_b$, and the CO LDMEs, respectively. The center values in Eqs. (\ref{uncertaity chib}) and (\ref{uncertaity upsilon}) are calculated at $m_{H^0}=125$ GeV, $m_{b}=4.9$ GeV, and $\mu_r=m_{H^0}/2$, with the LDMEs taken as the center values in Table 4 of Ref. \cite{Feng:2015wka}. To estimate the uncertainty, we vary $m_{H^0}$ in $\left[ 123 , 127 \right]$ GeV, $m_b$ in $\left[ 4.7 , 5.1 \right]$ GeV, $\mu_r$ in $\left[m_{H^0}/4,m_{H^0}\right]$ with $m_{H^0}=125$ GeV, and the LDMEs from the upper limit to the lower limit, respectively. The numerical results show that the ambiguities of $\mu_r$,  $m_{b}$, and the LDMEs are responsible for the main uncertainties, while varying $m_{H^0}$ only slightly influence the predictions on the total decay widths.

\section{Summary}

In this paper, we used NRQCD factorization to investigate the inclusive productions of the $\Upsilon(1S,2S,3S)$ and $\chi_b(1P,2P,3P)$ via the Standard Model Higgs boson decay up to $\mathcal O(\alpha\alpha_s^{2})$ order. It is found that the CO states, especially $ ^3S_1^{[8]} $, provide remarkable contributions, leading to vital effect on the predictions on the total decay widths. The newly calculated NLO QCD corrections to the lowest order process of $ ^3S_1^{[8]} $, $ H^0 \to b\bar{b}[^3S_1^{[8]}]+g $, can significantly (3-4 times) enhance the LO results, subsequently enlarging the total $ ^3S_1^{[8]} $ contributions by about $ 40\% $. In addition to the crucial effect on the total decay widths of $\Upsilon(nS)$ and $\chi_b(nP)$, including the CO states also influence the ratios of $ \frac{\Gamma_{\chi_{b2}}}{\Gamma_{\chi_{b0}}} $ and $ \frac{\Gamma_{\chi_{b2}}}{\Gamma_{\chi_{b1}}} $ a lot. Regarding the $ ^3P_{J}^{[1]} $ state, besides the dominant $ H^0 \to b\bar{b}[^3P_{J}^{[1]}] + b + \bar{b} $ process, the newly introduced light hadrons associated process, $ H^0 \to b\bar{b}[^3P_{J}^{[1]}] + g + g $, can also provide non-negligible contributions, especially for $ J=2 $. The feeddown contributions via the decay of the higher excited states are found to be substantial, significantly influencing the NRQCD predictions.  In the end, the branching ratios of $ H^0 \to \Upsilon(nS)+X $ and $ H^0 \to \chi_b(nP)+X $ are predicted to be on the order of $10^{-5}-10^{-4}$ and $10^{-6}-10^{-5}$, reflecting the great potential of these processes to be detected at high energy colliders. As a conclusion, the decay of Higgs boson into $\Upsilon(nS)$ and $\chi_b(nP)$ can be considered as an ideal laboratory not only to study the heavy quarkonium production mechanism, but also to understand the electroweak breaking mechanism especially the Yukawa couplings.

\section{Acknowledgments}
\noindent{\bf Acknowledgments}:
Z. Sun is supported in part by the Natural Science Foundation of China under the Grant No. 11647113. and No. 11705034., by the Project for Young Talents Growth of Guizhou Provincial
Department of Education under Grant No.KY[2017]135, and by the Project of GuiZhou Provincial Department of Science and Technology under Grant No. QKHJC[2019]1160. Y. Ma is supported by PITT PACC.\\

\end{document}